\documentclass[preprintnumbers,aps,preprint,nofootinbib]{revtex4-1}

\usepackage{booktabs}
\usepackage{footnote}

\usepackage{lipsum}
\usepackage{amsmath}    
\usepackage{amssymb}    
\usepackage{color}         
\usepackage{graphicx}     
\usepackage[T1]{fontenc}
\usepackage{times}         
\usepackage{mathrsfs} 
\usepackage{myterms}
\usepackage{hyperref} 
\hypersetup{
colorlinks=true,       
linkcolor=blue,        
citecolor=blue,        
filecolor=magenta,     
urlcolor=blue          
}
\usepackage[all]{hypcap} 

\usepackage[none]{hyphenat} 

\usepackage{color,xcolor,fancybox,epsf,rotating,colordvi}

\begin{document}

\preprint{\parbox{1.6in}{\noindent ~~~~~~}}

\title{Smuon contribution to muon $g \! -\! 2$ in Grand Unified supersymmetric theories}

\author{Weichao Li}
\affiliation{School of Physics and Technology, Wuhan University, Wuhan 430072, China}

\author{Haoxue Qiao}
\affiliation{School of Physics and Technology, Wuhan University, Wuhan 430072, China}

\author{Kun Wang}
\affiliation{College of Science, University of Shanghai for Science and Technology, Shanghai 200093, China}

\author{Jingya Zhu}
\email[]{zhujy@henu.edu.cn} 
\affiliation{School of Physics and Electronics, Henan University, Kaifeng 475004, China}

\date{\today}

\begin{abstract}
In GUT-scale constrained (GUTc) supersymmetric (SUSY) models, 
the mass of smuon $\tilde{\mu}_1$ is typically heavier than that of stau $\tilde{\tau}_1$, and stau co-annihilation is a typical annihilation mechanism of dark matter. 
However, light smuon is more favored by the muon $g-2$ anomaly, thus smuon-neutralino loop contribution to muon $g-2$ is usually smaller than that of sneutrino-chargino. 
Inspired by the latest muon $g-2$ results, we take the GUTc- Next-to-Minimal Supersymmetric Model (NMSSM) as an example, where the gaugino (Higgs) masses are not unified to the usual parameter $M_{1/2}$ ($M_0$), exploring its possibility of light smuon and its contribution to muon $g-2$. 
After complicated calculations and discussions, we conclude that in GUTc-NMSSM the smuon can be lighter than stau. 
In this light-smuon scenario, the contribution of smuon-neutralino loop to the muon $g-2$ can be larger than that of the sneutrino-chargino loop. 
The annihilation mechanisms of dark matter are dominated by multiple slepton or chargino co-annihilation. 
In our calculations, we consider also other latest related  constraints like Higgs data, SUSY searches, dark matter relic density and direct detections, etc. 
\end{abstract}


\maketitle

\section{Introduction}
\label{sec:intro}

The muon anomalous magnetic moment, also known as muon $g-2$ anomaly, has served as a sensitive probe of new physics for a long time \cite{Muong-2:2001kxu, Muong-2:2006rrc}.
It refers to the discrepancy of muon magnetic moment between model-independent experimental measurement and theoretical prediction in the Standard Model (SM), that is: 
\begin{eqnarray}
	\label{demuong-2}
	\Delta a_{\mu} = a_{\mu}^{\rm{ex}}-a_{\mu}^{\rm{SM}},
\end{eqnarray}
where the SM value \cite{Aoyama:2020ynm}
\begin{eqnarray}
	\label{SMmuong-2}
	a_{\mu}^{\rm{SM}} = 116~591~810(43) \times 10^{-11}. 
\end{eqnarray}
In August 2023, Fermilab released new measurement results for the muon magnetic moment,
\begin{eqnarray}
	\label{exmuong-2}
	a_{\mu}^{\rm{ex}} = 116~592~059(22) \times 10^{-11}\,,
\end{eqnarray}
with the deviation from SM value increasing from $4.2\sigma$ released in 2021 to the current $5\sigma$  \cite{Muong-2:2021ojo, Muong-2:2023cdq}.

Besides the muon $g-2$ anomaly, there are also other theoretical principles or experimental problems, such as naturalness, grand unification, and dark matter, implying new physics beyond the SM. 
Supersymmetry (SUSY) \cite{Fayet:1976et, Fayet:1977yc, Martin:1997ns} predicts a spin-different-by-half partner for each SM particle, thus can solve these problems naturally and draws a lot of attention from both theorists and experimentalists. 
After the muon $g-2$ 2023 result, several works appear attempting to interpret the anomaly in different SUSY models \cite{Khan:2023ryc, Dai:2023pli, Bisal:2023iip, Choudhury:2023yfg, Huang:2023zvs, Shafi:2023ksr, Goto:2023qch}. 

In SUSY models, it is known that additional smuon-neutralino, sneutrino-chargino, and Higgs-lepton loops can contribute to muon $g-2$, denoted as $a_{\tilde{\mu}}$, $a_{\tilde{\nu}}$, and $a_{H}$ respectively. 
Usually $|a_{H}|$ is much smaller than $|a_{\tilde{\mu}}|$ or $|a_{\tilde{\nu}}|$, for the small SM Yukawa couplings between Higgs and muon leptons. 
And in GUT-scale constrained (GUTc) superaymmetric (SUSY) models, it is usually that $|a_{\tilde{\mu}}| \lesssim |a_{\tilde{\nu}}|$ for smuon $\tilde{\mu}_1$ is typically heavy (heavier than the light stau $\tilde{\tau}_1$) and meanwhile a pair of charginos can be lighter \cite{Wang:2018vrr, Wang:2019biy, Wang:2020tap, Wang:2020dtb, Ma:2020mjz, Zhao:2022pnv, Li:2022etb}. 
So it is interesting to explore under current related constraints, whether the smuon-neutralino loop can contribute more to muon $g-2$ than the sneutrino-chargino loop, or $a_{\tilde{\mu}} > |a_{\tilde{\nu}}|$ in formula.

In this work, we study the unusual case that smuon $\tilde{\mu}_1$ lighter than stau $\tilde{\tau}_1$ in light of the 2023 muon $g-2$ result, and we do this in the GUTc- Next-to-Minimal Supersymmetric Model (NMSSM). 
The NMSSM extends the Minimal Supersymmetric Model (MSSM) by a singlet supermultiplet, thus has in addition a singlet-dominated Higgs scalar and a singlino-dominated neutralino, which results in interesting phenomenology of Higgs and dark matter, and also affect SUSY searches and muon $g-2$ \cite{Ellwanger:2023zjc, Bisal:2023mgz, Cao:2023juc, Dao:2023kzz, Binjonaid:2023rtc, Jia:2023xpx, Borah:2023zsb, Ellwanger:2009dp, Maniatis:2009re, Cao:2012fz, Ellwanger:2011aa, King:2012is, Ferrara:2010in, Belanger:2005kh, Gunion:2005rw}. 
We consider all the related constraints, particularly the latest 2023 or 2022 results of muon $g-2$ by Fermilab \cite{Muong-2:2023cdq} and dark matter direct search by LZ \cite{LZ:2022lsv} and XENONnT \cite{XENON:2023cxc}. 
To focus on the unusual light-smuon scenario we also require the smuon $\tilde{\mu}_1$ lighter than the stau $\tilde{\tau}_1$. 
Most parameters are input at the GUT scale, where the Higgs and gaugino masses are not required to unify as the well-known $M_0$ and $M_{1/2}$ parameters in the fully constrained SUSY models respectively. 

The structure of the remaining parts of this paper is outlined as follows. 
In Sec. \ref{sec:ana}, we provide an introduction to the smuon sector within the NMSSM and succinctly present the pertinent analytical equations.
Then Sec. \ref{sec:num} is dedicated to presenting the results of numerical calculations and discussions. 
Lastly, our primary conclusions are summarized in Sec. \ref{sec:conc}.

\section{The slepton sectors and muon g-2 contributions in NMSSM}
\label{sec:ana}

With an additional singlet superfield $\hat{S}$ and  discrete $Z_3$ symmetry, the NMSSM  has its superpotential $W_{\rm NMSSM}$ relate to that of MSSM $W_{\rm MSSM}$ by \cite{Ellwanger:2009dp, Maniatis:2009re}:
\begin{eqnarray}
	\label{nmssmW}
	W_{\rm NMSSM} = W_{\text{MSSM}}|_{\mu \to \lambda \hat{S}} + \frac{\kappa}{3}  \hat{S}^3, 
\end{eqnarray}
where $\lambda$ and $\kappa$ are two input parameters of NMSSM.
Thus the $\mu$ term is dynamically generated by the singlet when it gets vacuum expectation value (VEV) $\langle S \rangle$, i.e., $\mu_{\text{eff}} = \lambda \langle S \rangle$.

In SM the charged leptons can be left- and right-handed, while the neutral neutrinos can only be right-handed. 
So in NMSSM, there can be two charged sleptons and one neutral sneutrino in each generation. 
The mass matrix for the charged sleptons can be written as 
\begin{equation}\label{mumass}
M_{\tilde{\ell}_i}^2 \!\!=\!\! \begin{pmatrix}
	\!\! M_{L_i}^2 \!\!+\!\! \left(\sin^2\!\theta_W \!\!-\!\! \frac{1}{2}\right)\! m_Z^2\!\cos\!2\!\beta \!\!
	& 
	m_\mu (A_{E_i} \!\!-\!\! \mu\! \tan\!\beta) 
	\\
	m_\mu (A_{E_i} \!\!-\!\! \mu\! \tan\!\beta) 
	& 
	\!\! M_{E_i}^2 \!\!-\!\! m_Z^2\!\sin^2\!\theta_W\!\cos\! 2\!\beta \!\!
\end{pmatrix} \,,
\end{equation}
\noindent where $i=1,2,3$ is the generation index, $M_{L_i}$, $M_{E_i}$ are the soft masses of left- and right-handed sleptons respectively, and $A_{E_i}$ is the trilinear couplings in the slepton sector. 
For each generation, the left- and right-handed charged sleptons can mix to form two mass-eigenstate ones. 
To be convenient in this work, we also call the lighter one in the second (third) generation smuon (stau).

In NMSSM, the SUSY partner gauginos, Higgsinos, and singlinos with the same quantum numbers also mix into neutralinos and charginos. 
For the neutral sector, $\{\tilde{B}, \tilde{W}^0, \tilde{H}_u^0, \tilde{H}_d^0, \tilde{S}\}$ mix into five neutralinos $\{\tilde{\chi}_i^0\}$, with $i=1,...,5$, in mass-increasing order. 
For the charged sector, $\{\tilde{W}^\pm, \tilde{H}^\pm\}$ mix into two pairs of charginos $\{\tilde{\chi}_i^\pm\}$, with $i=1,2$, also in mass-increasing order. 
Thus the bino, wino, up- and down-type Higgsino, and singlino components in the $i$'th neutralino can be denoted as $|N_{i1}|^2$, $|N_{i2}|^2$, $|N_{i3}|^2$, $|N_{i4}|^2$, and $|N_{i5}|^2$ respectively. 
Similarly, the wino and Higgsino components in the $i$'th charginos can be denoted as $|C_{i1}|^2$ and $|C_{i2}|^2$ respectively. 
The lightest SUSY particle (LSP), the 1st neutralino $\tilde{\chi}^0_1$ in this work, served as dark matter. 

The contribution of smuon-neutralino loop to muon $g-2$, denoted as $a_{\tilde{\mu}}$, can be expressed as \cite{Martin:2001st}:
\begin{align}
	a_{\tilde{\mu}} &= \frac{m_\mu}{16\pi^2} \sum_{im} \bigg[ -\frac{m_\mu}{6m_{\tilde{\mu}_m}^2 (1-x_{im})^4} \left( |n^L_{im}|^2 + |n^R_{im}|^2 \right)  \nonumber \\
	&\quad \times (1-6x_{im}+3x_{im}^2 +2x_{im}^3 -6x_{im}^2 \text{ln}~x_{im}) \nonumber \\
	&\quad + \frac{m_{\tilde{\chi}_i^0}}{m_{\tilde{\mu}_m}^2 (1-x_{im})^3} \text{Re}\left( n^L_{im}n^R_{im} \right)(1-x_{im}^2 +2x_{im} \text{ln}~x_{im}) \bigg],
\end{align}
while that of the sneutrino-chargino loop, $a_{\tilde{\nu}}$, can be expressed as \cite{Martin:2001st}:
\begin{align}
	a_{\tilde{\nu}} &= \frac{m_\mu}{16\pi^2} \sum_{k} \left[ \frac{m_\mu}{6m_{\tilde{\nu}_\mu}^2(1-y_k)^4} \left( |c^L_k|^2 + |c^R_k|^2 \right) \right. \nonumber \\
	&\quad \times (2+3y_k -6y_k^2 +y_k^3 +6y_k\text{ln}~y_k) \nonumber \\
	&\quad \left. - \frac{m_{\tilde{\chi}_k^\pm}}{m_{\tilde{\nu}_\mu}^2(1-y_k^3)} \text{Re}\left( c^L_k c^R_k \right) (3-4y_k+y_k^2 +2\text{ln}~y_k) \right]. 
\end{align}
Here, $m_{\mu}$, $m_{\tilde{\mu}_m}$,  $m_{\tilde{\nu}_{\mu}}$, $m_{\tilde{\chi}_i^0}$ and $m_{\tilde{\chi}_k^{\pm}}$ represent the masses of muon, smuons, sneutrinos, neutralinos and charginos, respectively. 
And the mass ratios  $x_{im}=m_{\tilde\chi_i^0}^{2}/m_{\tilde\mu_m}^{2}$ and $y_k=m_{\tilde\chi_k^{\pm}}^{2}/m_{\tilde\nu_{\mu}}^{2}$ originate from the loop integrals, and the indices $i=1, ..., 5$, $k=1, 2$, and $m=1,2$ are from the neutralinos, chargino, and smuon, respectively. 
The variables $n_{im}^L$, $n_{im}^R$, $c_k^L$, and $c_k^R$ are related to the couplings and mixing matrixes in the neutralino, chargino, and smuon sectors, respectively.

\section{Numerical results and discussions}
\label{sec:num}

In this work, we first explore the parameter space  within the GUTc-NMSSM using the public code  $\textsf{NMSSMTools\_6.0.0}$ \cite{Ellwanger:2004xm, Ellwanger:2005dv, Ellwanger:2006rn}.
This exploration is conducted under a set of experimental and theoretical constraints to ensure the robustness and reliability of our results.
The parameter space we explored are: 
\begin{eqnarray}\label{NMSSM-scan}
	&& 0 \textless \lambda \textless 0.3,
	~~~|\kappa| \textless 0.7, \nonumber\\
	&&1\textless\tan\beta \textless 80,
	~~~|A_{0}|\textless 10 {\rm ~TeV} \,,\nonumber\\
	&&M_{0},|M_{1}|,|M_{2}|\textless 1 {\rm ~TeV}\,,\nonumber\\
	&&|M_{3}|,|A_{\lambda}|,|A_{\kappa}|,|\mu_{\rm eff}|\textless 20 {\rm ~TeV} \,,
\end{eqnarray}
where $M_0$ and $A_0$ represent the unified sfermion masses and trilinear couplings in the sfermion sector, and  $M_{1,2,3}$ denote the non-universal gaugino masses at the GUT scale. 
Calculations of the non-universal Higgs mass at GUT scale are carried out using minimization equations, with $\lambda$, $\kappa$, and $\mu_{\rm eff}\equiv \lambda \langle S \rangle$ as the input parameters at SUSY scale.

The constraints we impose in our analysis encompass the following aspects: 
\begin{itemize}
	\setlength{\itemsep}{0pt}
	\setlength{\parsep}{0pt}
	\setlength{\parskip}{0pt}
	\item[(i)] A SM-like Higgs boson of around 125 GeV, with signal strengths consistent with the latest data, evaluated by $\textsf{HiggsSignals-2.6.2}$ \cite{Bechtle:2020uwn, Bechtle:2013xfa}.
	\item[(ii)] Exclusion limits in searches for additional Higgs bosons at LEP, Tevatron, and LHC, imposed by $\textsf{HiggsBounds-5.10.2}$ \cite{Bechtle:2020pkv, Bechtle:2013wla, Bechtle:2008jh}. 
	\item[(iii)] Upper limits on dark matter relic density with uncertainty $\Omega h^2 \leq 0.131$  \cite{Tanabashi:2018oca, Hinshaw:2012aka, Ade:2013zuv}, and direct detections \cite{LZ:2022lsv, XENON:2023cxc}, which are calculated with \textsf{micrOMEGAs} \cite{Belanger:2006is, Belanger:2008sj} in \textsf{NMSSMTools}. 
	\item[(iv)] Exclusion limits from SUSY searches implemented in \textsf{SModelS-v2.3.2}  \cite{Kraml:2013mwa, Ambrogi:2018ujg, Khosa:2020zar, Alguero:2021dig}, including these of electroweakinos in multilepton channels \cite{CMS:2017moi, CMS:2018szt}, gluinos, and  first-two-generation squarks  \cite{ParticleDataGroup:2022pth}, etc. 
	\item[(v)] The combined experimental result of muon $g-2$ in 2023 \cite{Muong-2:2023cdq}. 
	\item[(vi)] Theoretical constraints, e.g., vacuum stability and the absence of a Landau pole below the GUT scale \cite{Ellwanger:2006rn}.
\end{itemize}
In addition, we also require smuon lighter than stau to focus on the light smuon scenario. 

\begin{figure*}[htbp]
	\centering
	\includegraphics[width=0.9\textwidth]{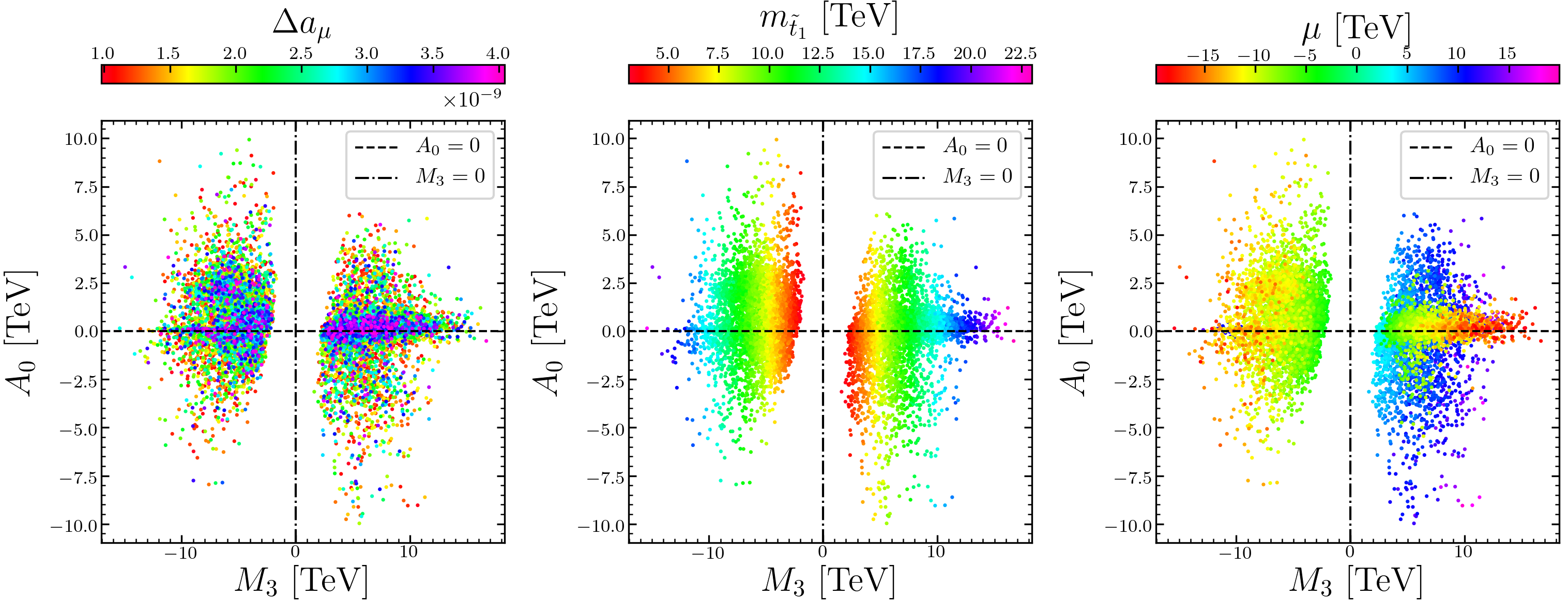}
	\vspace{-0.2cm}
	\caption{Surviving samples in the $A_0$ versus $M_3$ planes with colors indicating the SUSY contributions to muon g-2 $\Delta a_{\mu}$ (left), lighter stop mass $m_{\tilde{t}_1}$ (middle) and higgsino mass parameter $\mu$ (right), respectively.}
	\label{fig1}
\end{figure*}

\begin{figure*}[htbp]
	\centering
	\includegraphics[width=0.9\textwidth]{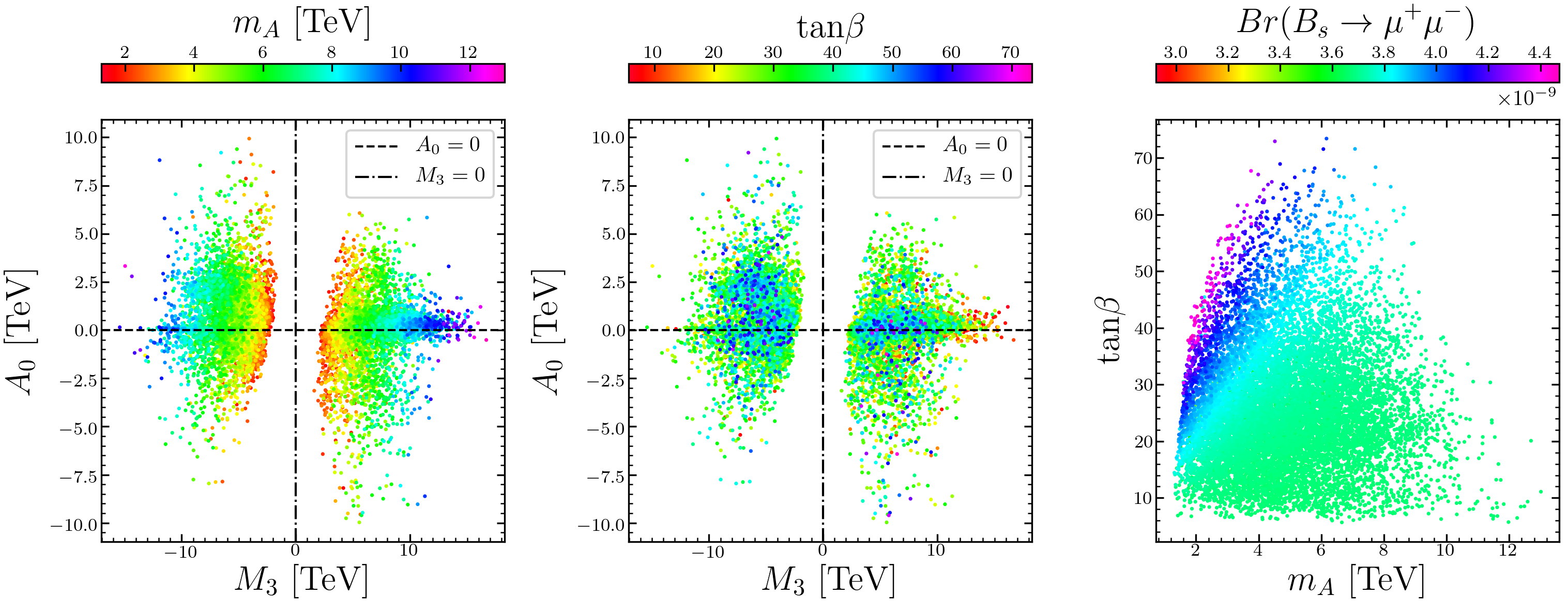}
	\vspace{-0.2cm}
	\caption{Surviving samples in the $A_0$ versus $M_3$ (left and middle), tan$\beta$ versus the CP-odd Higgs mass $m_A$ (right) planes. The colors indicate $m_A$ (left) planes, tan$\beta$ (middle), and $Br(B_s \to \mu^+\mu^-)$ (right), respectively.}
	\label{fig2}
\end{figure*}

\begin{figure*}[htbp]
	\centering
	\includegraphics[width=0.9\textwidth]{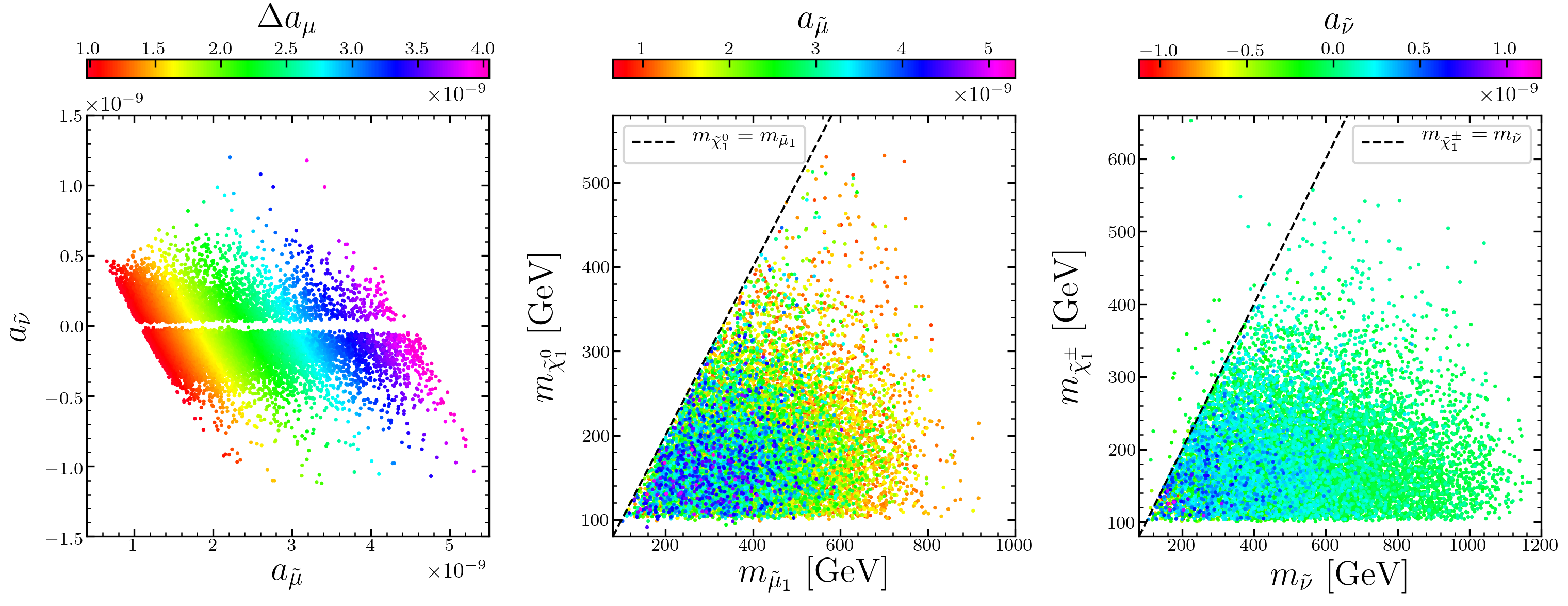}
	\vspace{-0.2cm}
	\caption{Surviving samples in the $a_{\tilde{\nu}}$ versus $a_{\tilde{\mu}}$ (left), $m_{\tilde\chi^0_1}$ versus $m_{\tilde\mu}$ (middle), and $m_{\tilde\chi^\pm_1}$ versus $m_{\tilde\nu}$ (right) planes, with colors indicating $\Delta a_{\mu}$ (left), $a_{\tilde{\mu}}$ (middle), and $a_{\tilde{\nu}}$ (right), respectively.}
	\label{fig3}
\end{figure*}

\begin{figure*}[htbp]
	\centering
	\includegraphics[width=0.9\textwidth]{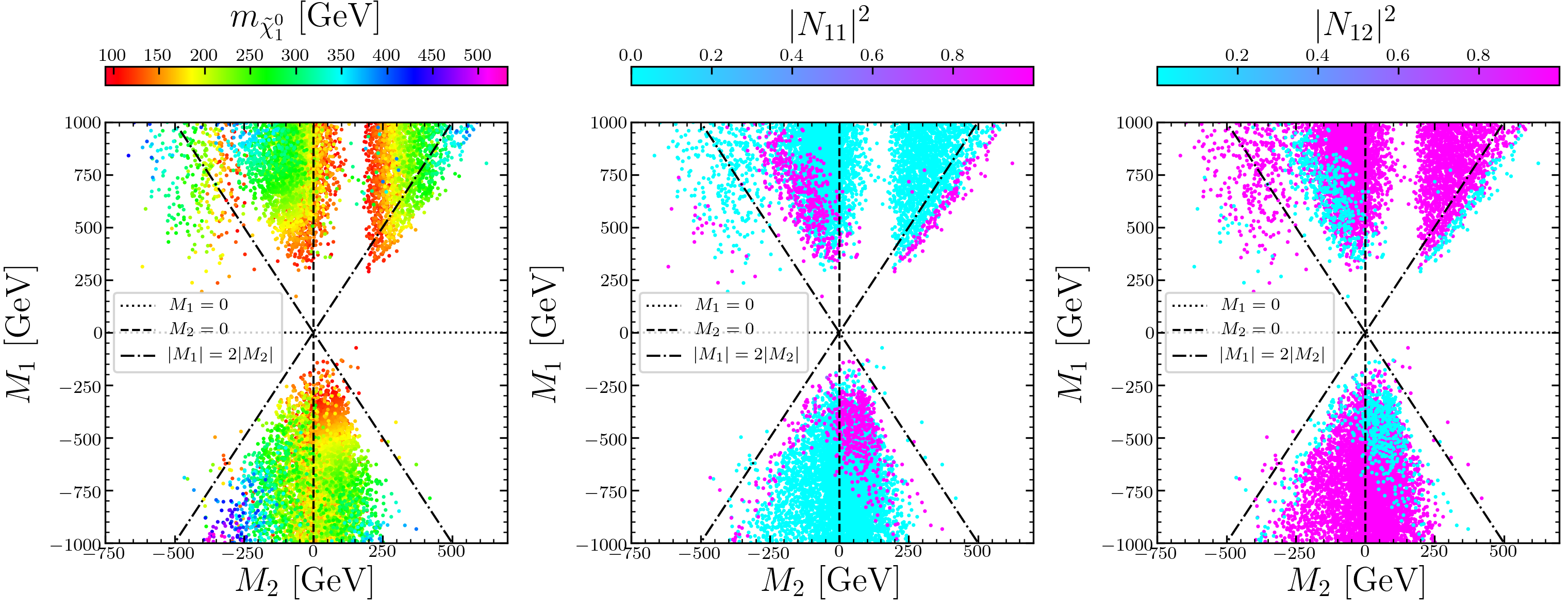}\\
	\includegraphics[width=0.9\textwidth]{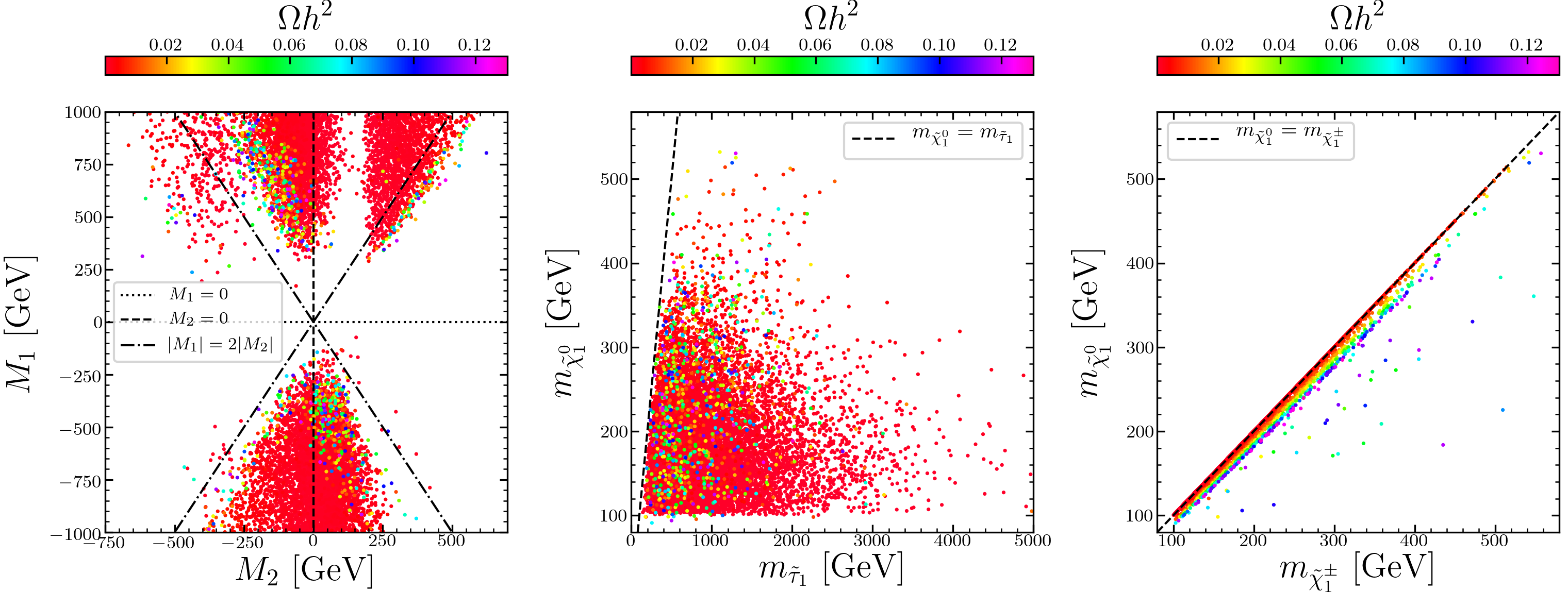}
	\vspace{-0.2cm}
	\caption{Surviving samples in the $M_1$ versus $M_2$ (upper 3 and lower left, where $M_{1,2}$ are both defined at GUT scale), LSP dark matter mass $m_{\tilde\chi_1^0}$ versus the lighter stau mass $m_{\tilde\tau_1}$ (lower middle), and $m_{\tilde\chi_1^0}$ versus the lighter chargino mass $m_{\tilde\chi_1^{\pm}}$ (lower right) planes. Colors indicate $m_{\tilde\chi_1^0}$ (upper left), bino component in LSP (upper middle), wino component (upper right), and LSP relic density ($\Omega h^2$), respectively.}
	\label{fig4}
\end{figure*}

\begin{figure*}[htbp]
	\centering
	\includegraphics[width=0.9\textwidth]{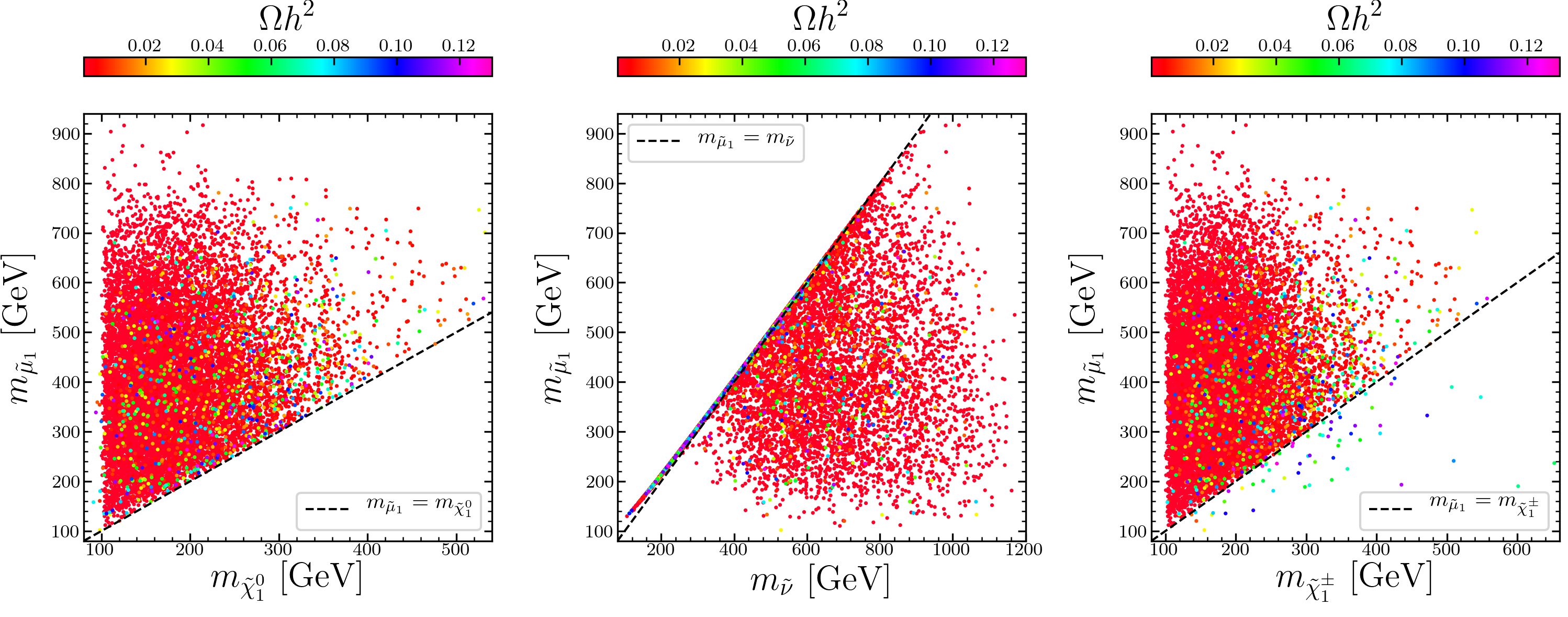}
	\vspace{-0.2cm}
	\caption{Surviving samples in the lighter smuon mass  $m_{\tilde\mu_1}$ versus LSP dark matter mass $m_{\tilde\chi_1^0}$ (left) and the lighter chargino mass $m_{\tilde\chi_1^{\pm}}$ (right) planes, with colors indicating LSP relic density $\Omega h^2$.}
	\label{fig5}
\end{figure*}

In Fig. \ref{fig1}, we project the surviving samples on the $A_0$ versus $M_3$ planes, with colors indicating $\Delta a_{\mu}$, the lighter stop mass $m_{\tilde{t}_1}$ (middle) and Higgsino mass parameter $\mu_{\rm eff}$ (right),  respectively. 
From this figure, one can see that the mass parameter of gluino $|M_3|$ and Higgsino $|\mu_{\rm eff}|$ should be large, which also lead to heavy stop ($m_{\tilde{t}_1}\gtrsim 1.5\TeV$) and non-SM doublet-dominated Higgs (with mass parameter $M_A\gtrsim 1.5\TeV$). 
The sign of $\mu_{\rm eff}$ is obviously not symmetric over that of $M_3$. 

Fig. \ref{fig2} shows the surviving samples on the $A_0$ versus $M_3$ (left and middle), and tan$\beta$ versus $m_A$ (right) planes, with colors indicating $m_A$ (left), $\tan\beta$ (middle), and $Br(B_s \to \mu^+\mu^-)$ (right), respectively.
From the left and middle planes, one can see that $\tan\beta\gg 1$, and the rare decay ratio $B_s\to\mu^+\mu^-$ is obviously related to $M_A$ and $\tan\beta$. 
From the right plane, one can see that the correlation between muon $g-2$ and $A_0$ versus $M_3$ is not obvious. 
But one can see that most large-$\Delta a_{\mu}$ samples are located in small-$A_0$ regions. 

Fig. \ref{fig3} shows the surviving samples in the $a_{\tilde{\nu}}$ versus $a_{\tilde{\mu}}$ (left), $m_{\tilde\chi^0_1}$ versus $m_{\tilde\mu}$ (middle), and $m_{\tilde\chi^\pm_1}$ versus $m_{\tilde\nu}$ (right) planes, with colors indicating $\Delta a_{\mu}$ (left), $a_{\tilde{\mu}}$ (middle), and $a_{\tilde{\nu}}$ (right), respectively.
From this figure, one can see that for most samples in the light-smuon scenario, SUSY contributions to muon $g-2$ mainly come from the smuon-neutralino loops. 
The smuon-neutralino contribution $a_{\tilde{\mu}}$ can be $2\sim 50 \times 10^{-10}$, while the sneutrino-chargino one $a_{\tilde{\nu}}$ can be positive or negative. 
There are obvious correlations between $a_{\tilde{\mu}}$ and the mass plane of smuon-neutralino, and also between  $a_{\tilde{\nu}}$ and the mass plane of sneutrino-chargino.
For most samples, the light chargino mass $m_{\tilde{\chi}^\pm_1}$ is smaller than the muon-sneutrino mass $m_{\tilde{\nu}_{\mu}}$. 
But there are also some samples with the opposing relation. 
We also checked that the light charginos are wino-dominated. 

In the first four plots in Fig. \ref{fig4}, surviving samples are projected on the planes of $M_1$ versus $M_2$, with color denoting mass, bino and wino components, and relic density (RD) of the lightest neutralino $\tilde\chi_1^0$, respectively. 
The last two plots show dark matter relic density on the mass planes of $m_{\tilde{\chi}^0_1}$ versus $m_{\tilde{\tau}_1}$ and $m_{\tilde{\chi}^\pm_1}$ respectively. 
From the upper plots, one can see that the lightest neutralino $\tilde{\chi}^0_1$, or LSP,  is bino- or wino-dominated. 
They also bear out the relation between gaugino masses at GUT and EW scale: when $\tilde{\chi}^0_1$ is bino-like, $m_{\tilde{\chi}^0_1}\approx 0.4 M_1$; when $\tilde{\chi}^0_1$ is wino-like, $m_{\tilde{\chi}^0_1}\approx 0.8 M_2$. 
Combined with the lower left plane, one can see that when RD is sufficient, the LSP $\tilde{\chi}^0_1$ should be bino-like. 
With mass lighter than about $500\GeV$, the wino-like LSP always gets small RD. 
From the last two plots, one can see all low-RD samples ($\Omega h^2 \lesssim 0.02$) are with $\tilde{\chi}^0_1$ and $\tilde{\chi}^\pm_1$ mass-degenerate, while there are some sufficient-RD samples without $\tilde{\chi}^0_1$ mass-degenerate $\tilde{\tau}_1$ or $\tilde{\chi}^\pm_1$. 
That means there are some annihilation mechanisms other than the usual stau or chargino co-annihilation. 

Since smuon is lighter than stau, we investigate the  annihilation mechanisms caused by light smuon. 
In Fig. \ref{fig5} we project surviving samples in smuon versus LSP (left), muon sneutrino (middle), and light chargino mass (right), with colors indicating LSP RD $\Omega h^2$. 
From the right plane, one can see that all small-RD samples are with the smuon heavier than light chargino $\tilde{\chi}^\pm_1$, while there are also sufficient-RD samples with smuon lighter than $\tilde{\chi}^\pm_1$. 
From the left plane, one can see that there are also sufficient-RD samples with mass-degenerate $\tilde{\mu}$ and LSP. 
Combined with the three planes, one can know that for sufficient-RD samples there can be complicated mult-slepton co-annihilation mechanisms caused by light smuon. 

\section{Conclusions}
\label{sec:conc}

In this work, in light of the latest muon $g-2$ result, we focus on light smuon in the GUT-scale constrained (GUTc) NMSSM. 
We first scan the parameter space, considering a series of related constraints, including muon $g-2$, Higgs data, non-SM Higgs and SUSY searches, dark matter relic density and detections, etc. 
We also require $m_{\tilde{\mu}_1} < m_{\tilde{\tau}_1}$ to ensure the unusual light-smuon scenario, for in the usual case of GUTc-SUSY models stau is heavier than smuon. 
Then with the surviving samples, we investigated  phenomenological aspects such as their parameter space, mass spectrum, muon $g-2$ contributions, dark matter annihilations, etc.
After complicated analysis, we get the following conclusions in GUTc-NMSSM: 
(i) Smuon can be lighter than stau with non-universal gaugino masses. 
(ii) The SUSY contributions to muon $g-2$ can be dominated by the smuon-neutralino loops with light smuon. 
(iii) The annihilation mechanisms of dark matter are dominated by multiple slepton or chargino co-annihilation.

\section*{Acknowledgments.}
This work was supported by the National Natural Science Foundation of China (NNSFC) under Grant Nos. 12275066, 11605123, and 12074295.


\end{document}